\documentclass[journal=acsnano,manuscript=article]{achemso}
\usepackage{chemformula} % Formula subscripts using \ch{}
\usepackage[utf8]{inputenc}
\usepackage{graphicx}
\usepackage{url}
\usepackage[normalem]{ulem}
\usepackage{color}
\usepackage{mathtools,amssymb}
\usepackage{textcomp}
\usepackage{gensymb}
\usepackage[draft]{todonotes}
\usepackage{braket}
\title{Photon assisted tunneling of high order multiple Andreev reflections in epitaxial nanowire Josephson junctions}

\author{Damon James Carrad}
\affiliation{%
Center for Quantum Devices, Niels Bohr Institute, University of Copenhagen, 2100 Copenhagen, Denmark}
\alsoaffiliation{%
Department of Energy Conversion and Storage, Technical University of Denmark, Fysikvej, Building 310, 2800 Kgs. Lyngby
}%
\altaffiliation{These two authors contributed equally}
\email{damonc@dtu.dk}

\author{Lukas Stampfer}
\affiliation{%
Center for Quantum Devices, Niels Bohr Institute, University of Copenhagen, 2100 Copenhagen, Denmark}
\altaffiliation{These two authors contributed equally}

\author{D\={a}gs Ol\u{s}teins}%
\affiliation{%
Center for Quantum Devices, Niels Bohr Institute, University of Copenhagen, 2100 Copenhagen, Denmark}

\author{Christian Emmanuel Noes Petersen}%
\affiliation{%
Center for Quantum Devices, Niels Bohr Institute, University of Copenhagen, 2100 Copenhagen, Denmark}

\author{Sabbir A. Khan}%
\affiliation{%
Center for Quantum Devices, Niels Bohr Institute, University of Copenhagen, 2100 Copenhagen, Denmark}
\alsoaffiliation{%
Danish National Metrology Institute, Kogle Alle 5, 2970 H{\o}rsholm, Denmark}

\author{Peter Krogstrup}%
\affiliation{%
Center for Quantum Devices, Niels Bohr Institute, University of Copenhagen, 2100 Copenhagen, Denmark}

\author{Thomas Sand Jespersen}%
\affiliation{%
Center for Quantum Devices, Niels Bohr Institute, University of Copenhagen, 2100 Copenhagen, Denmark}
\alsoaffiliation{%
Department of Energy Conversion and Storage, Technical University of Denmark, Fysikvej, Building 310, 2800 Kgs. Lyngby
}%
\email{tsaje@dtu.dk}

\begin{document}

\begin{abstract}
Semiconductor/superconductor hybrids exhibit a range of phenomena that can be exploited for the study of novel physics and the development of new technologies. Understanding the origin the energy spectrum of such hybrids is therefore a crucial goal. Here, we study Josephson junctions defined by shadow epitaxy on InAsSb/Al nanowires. The devices exhibit gate-tunable supercurrents at low temperatures and multiple Andreev reflections (MARs) at finite voltage bias. Under microwave irradiation, photon assisted tunneling (PAT) of MARs produces characteristic oscillating sidebands at quantized energies, which depend on MAR order, $n$, in agreement with a recently suggested modification of the classical Tien-Gordon equation. The scaling of the quantized energy spacings with microwave frequency provides independent confirmation of the effective charge $ne$ transferred by the $n^\mathrm{th}$ order tunnel process. The measurements suggest PAT as a powerful method for assigning the origin of low energy spectral features in hybrid Josephson devices.

\end{abstract}
\maketitle

%introduction&motivation
Traditional Josephson junctions (JJ)\cite{Josephson:1962} -- where two superconductors are coupled through a thin tunnel barrier -- constitute the most important device element in superconducting electronics and in state-of-the-art superconducting quantum information processors\cite{kjaergaard:2020}. Recently, there has been an increasing interest in merging the quantum properties of superconductors with the electrostatic control of semiconductor electronics. For example, semiconductor-based JJs allow electrostatic tunability of the Josephson coupling,\cite{Doh:2005} enabling gate-tunable superconducting qubits,\cite{Larsen:2015} and qubit implementations based on the Andreev bound states of few-channel JJs\cite{Hays:2018} or topologically protected low-energy states in one-dimensional proximity-coupled semiconductor/superconductor hybrid devices\cite{Oreg:2010,Lutchyn:2010}. In all cases, the electronic states in semiconductor/superconductor hybrid materials are of fundamental importance. The spectrum can be probed using normal/superconductor spectroscopy\cite{giaever:1960,deacon:2010,pillet:2010}, microwave spectroscopy\cite{ChauvinPRL2006} or by measuring finite-bias transport in semiconductor JJs\cite{heedt:2021}. In the latter case, however, low energy spectral features can both be related to states of the leads or to higher order transport processes\cite{Octavio:1983,Blonder:1982,Bratus:1995,AverinPRL1995}. Sharp spectral features are expected to appear at voltages $\pm 2\Delta/ne$ due to the sequential opening and closing of the allowed orders $n = 1,2,..$ of multiple Andreev reflection (MAR) processes. The MAR energies are independent of the junction transparency and analysis of the sub-gap spectrum allows extraction of the superconducting gap\cite{kjaergaard:2017, heedt:2021} and the determination of channel transmission of the junction\cite{Scheer:1998,goffman:2017}. In nanoscale semiconductor JJs, unambiguous assignment of MAR features in transport becomes problematic due to potential modifications of the sub-gap spectrum including from charging phenomena\cite{Buitelaar:2003, Eichler:2007, Sand-Jespersen:2007} and intrinsic properties of the semiconductor such as spin-orbit coupling\cite{dolcini:2008,hajer:2019}. This is increasingly important for processes of higher-orders -- i.e.\ lower energy -- where strong peaks unrelated to MAR can appear in the tunnel spectrum\cite{abay_charge_2014}, heating effects may modify the gap and thus the MAR positions, and MAR orders have been observed to be missing from the spectrum for reasons not yet understood\cite{xiang:2006}. Looking beyond the hybrid device studied in this manuscript, understanding of sub-gap spectra is crucial for interpreting results from the powerful tool of superconducting scanning tunneling microscopy,\cite{peters:2020} which allows special mapping of sub-gap states and is applicable to a wide variety of materials.

A characteristic feature which could distinguish MAR processes and their orders is the transferred charge: The $n^\mathrm{th}$-order process, dominating at $V_{sd}=2\Delta/ne$, transfers a charge $q_n = n e$ across the junction\cite{Cuevas:1999}. While the sub-gap spectrum has been extensively investigated\cite{Octavio:1983,Blonder:1982,Bratus:1995,AverinPRL1995}, only a few studies have directly addressed $q_n$. In Refs.\ \cite{Cron:2001,Kozhevnikov:2000,Ronen:2016} $q_n$ was investigated through the bias scaling of the noise spectrum of JJ which was found to be consistent with the prediction. Recently, an alternative method for measuring $q_n$ has been suggested based on the evolution of the sub-gap spectrum under microwave irradiation\cite{kot:2020,roychowdhury:2015,peters:2020}. Due to the Tien-Gordon effect\cite{Tien:2963, Platero:2004} -- or photon assisted tunneling (PAT) -- modulating sidebands are expected in the non-linear sub-gap spectrum around the $n^\mathrm{th}$ MAR structure. While the conventional Tien-Gordon model predicts a characteristic modulation scale of $\delta V = \hbar f /e$ where $f$ is the frequency of the microwave field, Ref.\ \cite{roychowdhury:2015,kot:2020} suggested that in higher order processes the total transferred charge should be used. This was consistent with measurement of quasi-particle tunneling $(1e)$ and incoherent Cooper pair tunneling ($2e$), and is consistent with early studies of the sub-gap spectrum at fixed microwave frequency and power\cite{ChauvinPRL2006}. In this letter we present measurements at varying microwave power and frequency of PAT of the sub-gap spectrum in epitaxial semiconductor/superconductor InAsSb/Al nanowire JJs realized by shadow lithography\cite{khan_highly_2020, Carrad:2020}. The measurements show clear signatures of the charge $q_n = ne$ of the $n^\mathrm{th}$ order MAR up to $n=3$.

\begin{figure*}[t]
\includegraphics[width=18cm]{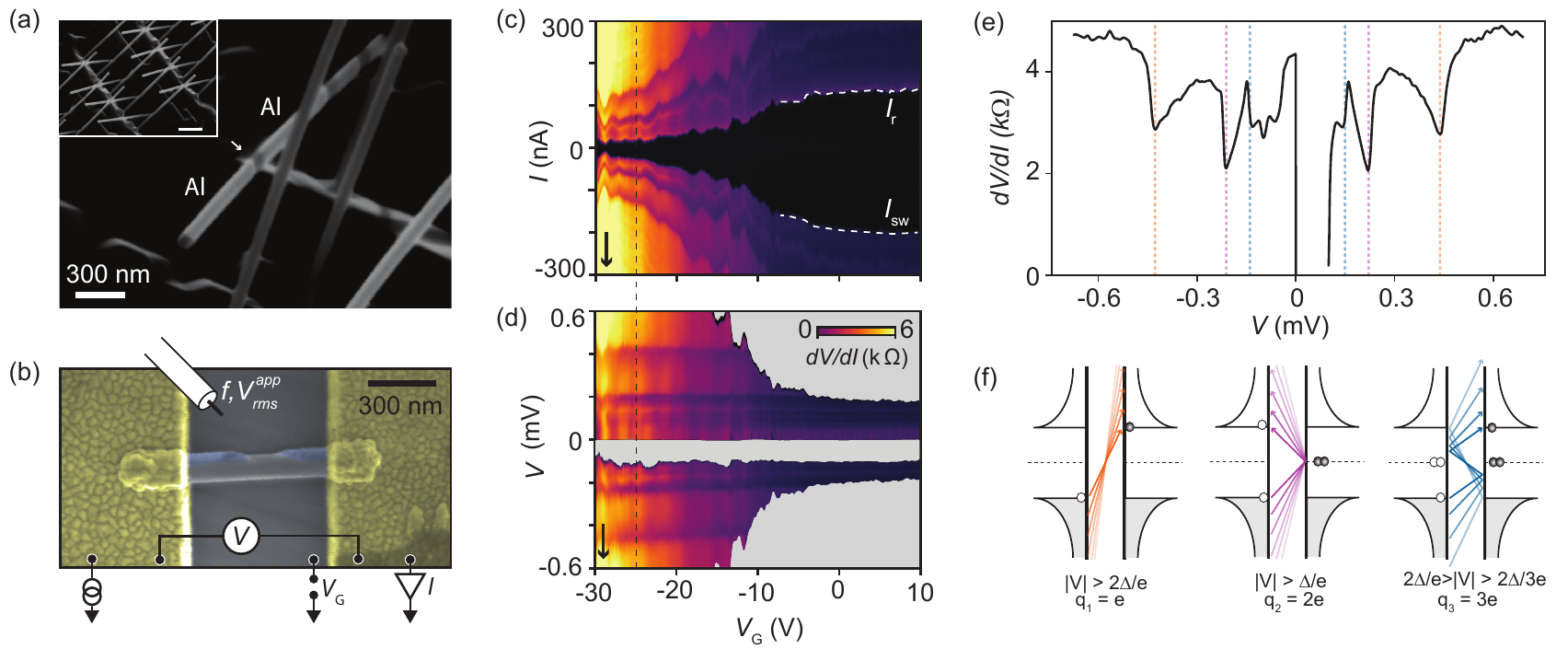}
\caption{\label{fig:Fig1} (a) Scanning electron micrograph (SEM) of as-grown InAsSb/Al nanowires. Al deposition was performed from roughly bottom-right to top-left across the image, and forward-positioned nanowires shadow those behind resulting in Al gaps (arrow). Scale bar in inset is $1 \, \mu \mathrm{m}$. (b) False colour SEM of a device featuring InAsSb nanowire (grey), shadow-patterned Al layer (blue) and Ti/Au ohmic contacts (gold). (c,d) Gate dependence at $T= 15 \, \mathrm{mK}$ 
of the differential resistance $\mathrm d V /\mathrm d I$ vs.\ measured current $I$, and voltage $V$, respectively. A zero resistance supercurrent is present between switching and retrapping currents $I_\mathrm{sw}$, $I_\mathrm{r}$. (e) Line trace in (d) at fixed $V_\mathrm G = -25 \, \mathrm V$. Dashed lines indicate features at voltages $\pm 2\Delta/e$ (orange), $\pm \Delta/e$ (purple), and $\pm 2\Delta/3e$ (blue) with $\delta = 225 \, \mu \mathrm{eV}$ attributed to $n = 1,2,3$ MAR. The corresponding transport processes are schematically illustrated in (f).}
\end{figure*}

InAsSb nanowires (NWs) were grown by molecular beam epitaxy following the procedure of Ref.\ \cite{khan_highly_2020}. Parallel trenches were etched in a $(100)$ InAs substrate and $90-120 \, \mathrm{nm}$ diameter NWs were grown from Au catalyst particles positioned on opposing $(111)$B inclined side facets. As shown in Fig.\ \ref{fig:Fig1}a, the growth position of the NWs were laterally offset such that neighboring NWs defined $\sim 100 \, \mathrm{nm}$ wide shadow gaps\cite{Gazibegovic:2017,Krizek:2017} in the $15 \, \mathrm{nm}$ thick, epitaxially matched Al film (arrow) \cite{Krogstrup:2015, khan_highly_2020, Carrad:2020, sestoft:2018}. NWs were transferred to a doped silicon substrate capped with $100 \, \mathrm{nm}$ SiO$_2$, and contacts to the Al half-shell were defined using established lithographic methods\cite{Krogstrup:2015, Chang:2015, Carrad:2020, khan_highly_2020}. Figure \ref{fig:Fig1}b shows a SEM micrograph of a typical finished device and schematic measurement circuit. The conducting Si substrate acted as back-gate, which was used to alter the carrier density in the exposed InAsSb. Each contact electrode was split into two individual bond pads to facilitate pseudo-four terminal measurements of the device resistance, eliminating series resistances from e.g. low pass filters in the cryostat cabling. The remaining series resistance -- likely dominated by the contacts --  $\sim 3 \, \mathrm k \Omega$, was identified as a constant off-set in the supercurrent regime and was subtracted in the following. Measurements were performed in a dilution refrigerator with a base temperature of $15 \, \mathrm{mK}$ and microwave radiation was coupled to the device by a coaxial line with $\sim 12 \, \mathrm{mm}$ of the inner conductor left exposed and situated $\sim 5 \, \mathrm{mm}$ above the sample.

Figure \ref{fig:Fig1}c shows the differential resistance, $dV/dI$, as a function of current through the device, $I$, and back-gate voltage $V_\mathrm G$. A zero resistance supercurrent is observed for low $|I|$. The current was swept from positive to negative (arrow) and the switching(re-trapping) current where the device transitions to(from) a resistive state from(to) the superconducting state is indicated. The differences in $I_\mathrm{sw}$ and $I_\mathrm{r}$ are commonly observed in nanowire JJs\cite{Doh:2005} and ascribed to a consequence of Joule heating in the dissipative regime and/or underdamped dynamics of the JJ\cite{mccumber_effect_1968,tinkham_hysteretic_2003}. Both $I_\mathrm{sw}$ and $I_\mathrm{r}$ decrease upon lowering $V_\mathrm G$ as expected from a JJ with constant $I_\mathrm c R_\mathrm N = \pi \Delta/2e$ and an $n$-type semiconductor which depletes with lowering $V_\mathrm G$. Here $I_\mathrm{c}$ is the critical current and $R_\mathrm{N}$ is the normal-state device resistance. The aperiodic oscillations in $I_\mathrm{sw}(V_\mathrm G)$ appearing in addition to the overall gate-dependence is attributed to universal conductance fluctuations in $R_\mathrm{N}(V_\mathrm G)$ \cite{Doh:2005}. At $V_\mathrm G \sim -25 \, \mathrm V$ the normal state resistance is $R_\mathrm{N} \sim 4.5\mathrm{k}\Omega$ and $I_\mathrm{sw} R_\mathrm N = 200~\mathrm{\mu V}$, relatively close to the theoretical value $\pi \Delta_\mathrm{Al}/(2e) = 350~\mathrm{\mu V}$, as previously reported for similar hybrids\cite{khan_highly_2020}. The measured excess current (Supporting Fig.\ S1) $I_\mathrm{ex} \sim 50 \, \mathrm{nA}$ corresponds to a junction transmission of $\sim 0.7$\cite{flensberg_subharmonic_1988}. 

For $I$ above/below the zero-resistance region in Fig.\ \ref{fig:Fig1}c, $dV/dI$ exhibits a rich structure which follows the trend of $I_\mathrm{sw}$ and $I_\mathrm r$. To gain insight into the origin of these features, Fig.\ \ref{fig:Fig1}d shows the same data as a function of the measured voltage $V$. Upon sweeping $I$ down from zero, the voltage remains $V=0 \, \mathrm{V}$ until it abruptly switches at $I_\mathrm{sw}$ to a finite value. Therefore no data is obtained with $V$ below this value, and this region is greyed-out in Fig.\ \ref{fig:Fig1}d. Conversely, $V$ gradually decreases to zero around $I_\mathrm{r}$ when sweeping $I$ down from positive values due to the underdamped junction dynamics,\cite{mccumber_effect_1968,tinkham_hysteretic_2003} enabling data extraction down to $V=0$ for $I>0$. As expected for MAR, the structure in Fig.\ \ref{fig:Fig1}d appears at fixed voltages despite the variation in conductance with $V_\mathrm G$. Figure \ref{fig:Fig1}e shows a line trace of $dV/dI$ vs.\ $V$ for fixed $V_\mathrm G = -25 \, \mathrm V$. The three most pronounced dips are highlighted by dashed lines and follow $\pm 2\Delta/e$ (orange), $\pm \Delta/e$ (purple), and $\pm 2\Delta/3e$ (blue) with $\Delta = 225 \, \mu \mathrm{eV}$ consistent with prior results from 15~nm thick Al \cite{deng_majorana_2016, Chang:2015, khan_highly_2020}. This series is expected for MAR and the relevant processes transfer charges of $1e, 2e$ and $3e$ and are illustrated in Fig.\ \ref{fig:Fig1}f, where the slopes of the arrows represent voltage biases. Direct quasi particle tunneling $(1e)$ occurs for $|V| \ge 2\Delta/e$. In the general case of $n \ge 2$, the $n^\mathrm{th}$ order process involves $n-1$ Andreev reflections, a charge transfer of $ne$, and is allowed for $|V| \ge 2\Delta/(ne)$. Suppression of $n \ge 3$ processes occur for $e|V| \ge 2\Delta /(n-2)$ due to the low probability of Andreev reflection above the gap\cite{Klapwijk:1982}. Overall, the dc transport characteristics in Fig.\ \ref{fig:Fig1} are comparable to previous NW-based JJs\cite{Doh:2005,abay_charge_2014,ridderbos_multiple_2019,xiang:2006}.

\begin{figure*}[t]
\includegraphics[width=18cm]{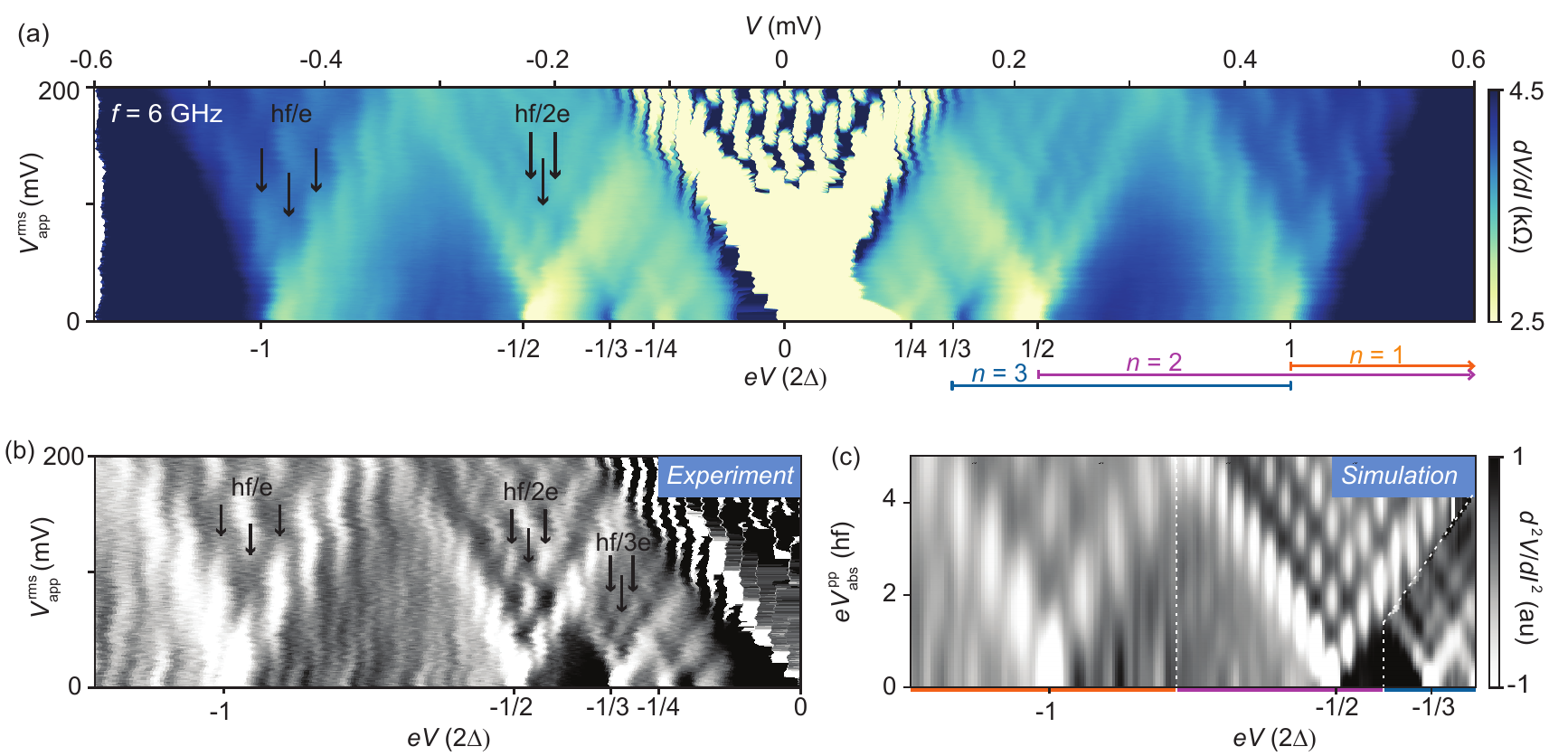}
\caption{\label{fig:Fig2} (a) Differential resistance at $V_\mathrm{G} = -25 \mathrm V$ as a function of bias, $V$, and the rms amplitude of $6\,\mathrm{GHz}$ microwave radiation. The bottom axis represents energy $E$ in units of $2\Delta$, with $\Delta = 225~\mu$eV and the energy intervals for the first three MAR processes are indicated by horizontal lines. (b) The second derivative, $d^2V/dI^2$, in a region of (a) around the higher order MAR showing the Tien-Gorden modulation in the $n=2$ and $n=3$ MAR processes. The changing period of modulation is indicated by arrows. (c) A simulation of the data in (b), based on Eq.\ 1 modified to account for the charge of the higher order transport processes. Processes of different order are assumed to dominate in the indicated regions.}
\end{figure*}

We now consider the features arising from irradiating the device with microwaves. Figure \ref{fig:Fig2}a shows $dV/dI$ vs.\ $V$ and the rms amplitude, $V_\mathrm{app}^\mathrm{rms}$, of $6 \, \mathrm{GHz}$ microwave radiation, as applied to the antenna (Fig.\ \ref{fig:Fig1}b). The gate voltage was fixed at $V_\mathrm{G}=-25 \mathrm{V}$ as in Fig.\ \ref{fig:Fig1}e -- similar features were observed at other $V_\mathrm{G}$, see Supporting Section II -- and $I$ was swept from negative to positive. A series of `V'-shaped patterns are observed, fanning out from $V_\mathrm{app}^\mathrm{rms} = 0$. The structure emerging from $V=0\,\mathrm V$ -- i.e., when the JJ carries a supercurrent -- is due to Shapiro steps\cite{Shapiro:1963, Josephson:1962} caused by phase-locking between the microwave field and the superconducting phase difference across the JJ leading to at step-wise increase of $V$ in units of $hf/(2e)$. In Supporting Fig.\ S3, we replot this regime as a function $I$ as is common for studies of Shapiro steps, showing the conventional Shapiro pattern\cite{Doh:2005, Larson:2020, ridderbos_multiple_2019,perla:2021}. The measurements are well-described by the extended RCSJ model and will not be considered further here.

The remaining V-shaped structures in Fig.\ \ref{fig:Fig2}a, which evolve from the MAR features at finite $V$, have a different origin and are the main focus of this work. Since the junction does not carry a supercurrent, phase locking and the resultant Shapiro steps do not occur. Instead, we attribute the structures to photon assisted tunneling. In general, for any two terminal device absorbing radiation at frequency $f$, charges can absorb/emit photons and thereby gain energy $\pm lhf$ and tunnel across the junction -- here $h$ is the Planck constant and $l = 0,\pm 1, \pm 2, \ldots$. Thus, at finite irradiation amplitude, the current at the dc voltage $V$ is related to the dc current in the absence of irradiation, $I^0$, at $V \pm hfl/e$. For smooth $I^0(V)$, this has no consequence, however, non-linearities in $I^0(V)$ will repeat at finite radiation amplitude as sidebands with an intensity modulated by the radiation amplitude. In the simplest case where the radiation field on the sample is sinusoidal, $V_\mathrm{abs} = V_\mathrm{abs}^\mathrm{pp} \cdot \sin (2\pi f t)$, the current at finite radiation amplitude is given by the Tien-Gordon equation \cite{Tien:2963}: 

\begin{equation}
I_\mathrm{dc}(V) = \sum_l J^2_l \left( \frac{e V_\mathrm{abs}^\mathrm{pp}}{hf} \right) \cdot I^0 \left(V + \frac{hfl}{e} \right)
\end{equation}

where $J_l(x)$ is the $l^\mathrm{th}$ Bessel function. The amplitude of each sideband oscillates with increasing $V_\mathrm{abs}^\mathrm{pp}$ according to $J_l^2$ (Supporting Fig.\ S4). This explains the overall shape of the structures in Fig.\ \ref{fig:Fig2}a which evolve from the non-linearities at $V=2\Delta/ne$ associated with MAR. The highly asymmetric MAR features in Fig.\ \ref{fig:Fig1}e makes the inflection points of $dV/dI$ -- i.e. peaks in $d^2V/dI^2$ -- well suited for identifying the onset of MAR as shown in Fig.\ \ref{fig:Fig2}b. The modulation period of the features associated with the $n=1,2,3$ order MARs can be clearly resolved and as highlighted by the black arrows, the modulation period is not the same for each `V'-fan but rather decreases with decreasing bias. This is inconsistent with the conventional Tien-Gordon model described above which predicts the same characteristic pattern evolving from all non-linearities in $I^0$. Instead, as discussed in the introduction, the pattern in Fig.\ \ref{fig:Fig2}a,b follows the work of Refs\ \cite{kot:2020,roychowdhury:2015} suggesting that Eq.\ 1 describes the microwave response for higher-order tunnel processes if the electron charge, $e$, is replaced by the total transferred charge $q=ne$: Refs\ \cite{kot:2020,roychowdhury:2015} experimentally confirmed this for quasiparticle ($e$) and incoherent Cooper pair tunneling at zero bias ($2e$). For our case, the $n^\mathrm{th}$ order MAR transfers $ne$ electrons (Fig.\ \ref{fig:Fig1}f), and the corresponding modulation period should be $\Delta V_n = hf/ne$. For $f=6 \, \mathrm{GHz}$ this amounts to $24.8\, \mu \mathrm{V}$, $12.4\, \mu \mathrm{V}$, and $8.3\, \mu \mathrm{V}$ for $n=1,2,3$, respectively, in good agreement with the measured splittings $23.8\, \mu \mathrm{V}$,$12.2\, \mu \mathrm{V}$, and $9.3\, \mu \mathrm{V}$ from Fig.\ \ref{fig:Fig2}a,b (see below). Using this assumption we show in Fig.\ \ref{fig:Fig2}c the result of a simulation of $d^2I/dV^2$ taking the $I-V$ curve in absence of radiation as input and evolving according to Eq.\ 1 under the simplifying assumption that the $n^\mathrm{th}$ order MAR dominate in the region around $2\Delta/ne$ as shown by the dashed lines Fig.\ \ref{fig:Fig2}c (see Supporting Section IV for details). Comparing to Fig.\ \ref{fig:Fig2}b, the simulation accurately captures the main features of the measurement.

\begin{figure}[t]
\includegraphics[width=8.5cm]{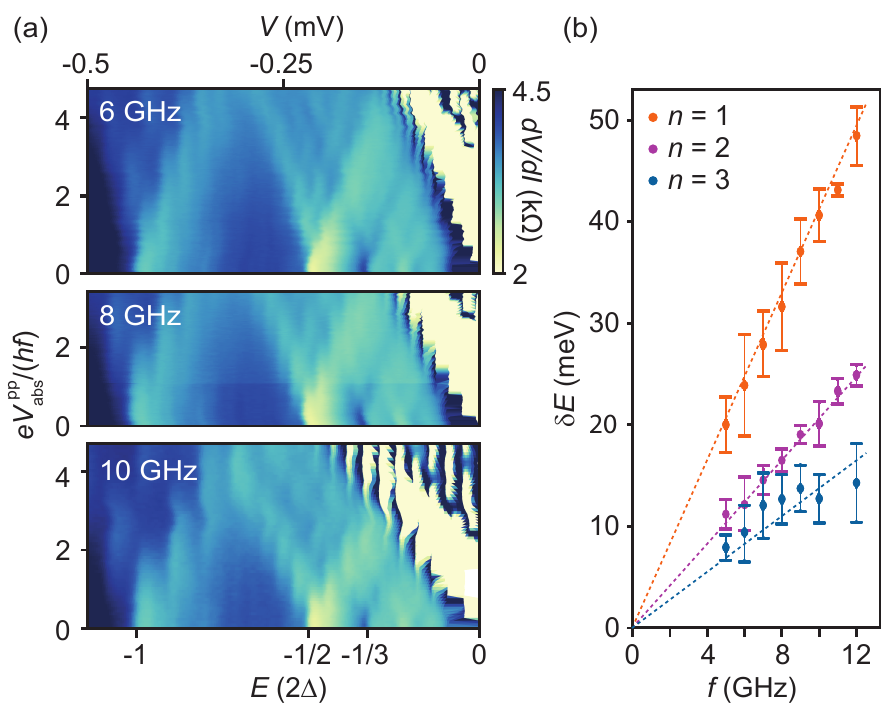}
\caption{\label{fig:Fig3} (a) $dV/dI$ vs $V$ and normalised absorbed microwave voltage $eV^\mathrm{pp}_\mathrm{abs}/(hf)$ for $f=6$,8 and 10~GHz. The 6~GHz data is a subset of Fig.~2a. The sideband spacing within each MAR order increases with increasing $f$. (b) Average sideband spacing $\left < \delta E \right >$ as a function of $f$ for $n=1,2,3$ (blue, orange and green data points). The blue, orange and green dashed lines show the theoretical $\delta E$ vs $f$ relationships for 1, 2 and 3 electron processes.}
\end{figure}

To further quantify the splitting, the measurements of Fig.\ \ref{fig:Fig2}a were repeated for different frequencies $f = 5 - 12$~GHz. Three representative examples are shown in Fig.\ 3a for $f = 6,8,10$~GHz. The coupling between the antenna and device depends strongly and non-monotonically on $f$, and therefore the amplitude of absorbed radiation, $V_\mathrm{abs}^\mathrm{pp}$ in Fig.\ \ref{fig:Fig3}a was for visualization purposes normalized to the first zero of $J_0$ which occurs at $eV_\mathrm{abs}^{pp}/(hf) = 2.4$ (see Supporting information for non-normalized data at all eight frequencies). Figure 3b shows the average of the sideband spacing $\left < \delta E \right > = e \left < \delta V \right >$, as a function of $f$ (data points) and the theoretical relationship $\delta E = hf/n$ (dashed lines) for $n = 1,2$ and 3 (blue, orange and green). The average was obtained across all the measured $\delta E$ within each order, and the error bars represent the standard deviation. The good agreement between theory and experiment quantitatively confirms that the $n^\mathrm{th}$ order MAR process involves the transmission of $n$ electrons across the junction. 

In conclusion, we have studied the finite-bias sub-gap structure in epitaxial InAsSb/Al JJs under microwave radiation. We showed that the microwave response provides a sensitive probe of the transmitted charge $q = ne$ up to order $n=3$ -- i.e.\ a MAR process including two Andreev reflections. The modulation period was found to decrease with increasing MAR order, and the dependence of the sideband spacings on the radiation amplitude agrees quantitatively with the expectation that each order transmits $n$ electrons across the junction. A simulation based on a modified version of the Tien-Gordon equation was found to accurately reproduce the measurements. We expect the effect to be observable in non-epitaxially matched hybrids, albeit likely with less sharp features than observed here,\cite{Chang:2015} and possibly modified weight of the MAR due to changes in the interface transmission.\cite{AverinPRL1995} We believe that the results will be valuable tool for understanding the low-energy sub-gap spectrum in hybrid nanoscale devices. This becomes increasingly relevant with the rapid developments in the quality of hybrid quantum materials, device fabrication and filtering and measurement techniques, which allows for increasing coherence lengths and increased spectral resolution and thus even higher order MAR at progressively lower energies. Finally, the identification of sub-gap states is a crucial task for the powerful technique of spatially resolved superconducting scanning tunneling microscopy.

\textbf{Acknowledgement}
This research was supported by the Danish National Research Foundation, Microsoft Quantum Materials Lab, and by research grants from Villum Fonden (00013157), The Danish Council for Independent Research (7014-00132), and European Research Council (866158).

\section*{Competing interests} The authors declare no competing interests.

\section*{Supporting Information} The supporting information contains additional data relating to the excess current, Shapiro steps, PAT of MAR at a different gate voltage, the underlying measurements for Fig. 3, as well as extra information regarding the simulation in Fig. 2c. This material is available free of charge via the internet at http://pubs.acs.org.

%\bibliography{refs,Patrefs,TSJzotero}
\providecommand{\latin}[1]{#1}
\makeatletter
\providecommand{\doi}
  {\begingroup\let\do\@makeother\dospecials
  \catcode`\{=1 \catcode`\}=2 \doi@aux}
\providecommand{\doi@aux}[1]{\endgroup\texttt{#1}}
\makeatother
\providecommand*\mcitethebibliography{\thebibliography}
\csname @ifundefined\endcsname{endmcitethebibliography}
  {\let\endmcitethebibliography\endthebibliography}{}

\end{document}